\documentclass[12pt,preprint]{aastex}

\begin{document}



\title{On MHD jet production in the collapsing and rotating envelope}

\author{Daniel Proga,\altaffilmark{1}}

\affil{$^1$ Princeton University Observatory, Peyton Hall, Princeton, NJ 08544.e-mail: dproga@astro.princeton.edu}

\def\LSUN{\rm L_{\odot}}
\def\MSUN{\rm M_{\odot}}
\def\RSUN{\rm R_{\odot}} 
\def\MSUNYR{\rm M_{\odot}\,yr^{-1}}
\def\MSUNS{\rm M_{\odot}\,s^{-1}}
\def\MDOT{\dot{M}}

\newbox\grsign \setbox\grsign=\hbox{$>$} \newdimen\grdimen \grdimen=\ht\grsign
\newbox\simlessbox \newbox\simgreatbox
\setbox\simgreatbox=\hbox{\raise.5ex\hbox{$>$}\llap
     {\lower.5ex\hbox{$\sim$}}}\ht1=\grdimen\dp1=0pt
\setbox\simlessbox=\hbox{\raise.5ex\hbox{$<$}\llap
     {\lower.5ex\hbox{$\sim$}}}\ht2=\grdimen\dp2=0pt
\def\simgreat{\mathrel{\copy\simgreatbox}}
\def\simless{\mathrel{\copy\simlessbox}}

\begin{abstract}
We present results from axisymmetric, time-dependent 
hydrodynamical (HD) and magnetohydrodynamical
(MHD) simulations of a gaseous envelope collapsing onto a  black hole (BH). 
We consider gas with so small angular momentum that after 
an initial transient, the flow in the HD case, accretes directly onto a BH
without forming a rotationally support torus.
However, in the MHD case even with a very weak initial magnetic field, 
the flow settles into a configuration with four components: 
(i) an equatorial inflow,
(ii) a bipolar outflow, (iii) polar funnel outflow, and
(iv) polar funnel inflow.
We focus our analysis on the second flow component of the MHD flow
which represents a simple yet robust example 
of a well-organized inflow/outflow solution to the problem
of MHD jet formation.
The jet is heavy, highly magnetized, and 
driven by magnetic and centrifugal forces. 
A significant fraction of the total energy in the
jet is carried out by a large scale magnetic field.
We review previous simulations, where
specific angular momentum was higher than that assumed here, 
and conclude that our bipolar outflow develops
for a wide range of the properties of the flow
near the equator
and near the poles. Future work
on such a simple inflow/outflow solution
will help to pinpoint the key elements
of real jets/outflows as well as help to 
interpret much more complex simulations aimed at 
studying jet formation and collapse of magnetized envelopes.
\end{abstract}


\section{Introduction}
The common occurrence and importance of astrophysical jets
have stimulated many theoretical studies.
Both analytic and numerical studies 
provide strong support for the scenario where jets are 
magnetohydrodynamic (MHD) in character and are associated
with an accretion disk, a rotating accretor, or both
(e.g., Blandford \& Payne 1982; Blandford 1990;
De Villier et al. 2003). In fact,
most MHD simulations of accretion flows show outflows
(e.g., Uchida \& Shibata 1985; Stone \& Norman 1994;
Hawley \& Balbus 2002; De Villier et al. 2004; Mizuno et al. 2004; 
Kato et al. 2004; McKinney \& Gammie 2004).

Here, we describe a study of the time evolution of
MHD flows in the vicinity of a stationary black hole (BH). This study
has been motivated by the results from our previous work 
(Proga \& Begelman 2003b, hereafter PB03b; Proga et al 2003) 
which showed that large-scale magnetic fields can produce
a jet from a rotationally supported accretion disk or torus
and also from extremely low angular momentum gas 
that almost radially accretes onto a BH.
This falling gas and associated outflow can be well-organized and
a simple, self-consistent solution for the MHD jet problem.
However, in our previous simulations and likely in some other simulations, 
this simple inflow/outflow is a component of a complex convolution
of a rotationally supported,
MHD turbulent torus, the torus corona and outflow.
To  articulate the basic physics that occurs in jet
production, we focus on a flow with angular momentum so low
that, if not for the effects of MHD, 
the flow would accrete directly onto a BH without forming a disk.

\section{Method}

To calculate the flow structure and evolution, we solve the equations of 
ideal MHD:

\begin{equation}
   \frac{D\rho}{Dt} + \rho \nabla \cdot {\bf v} = 0,
\end{equation}

\begin{equation}
   \rho \frac{D{\bf v}}{Dt} = - \nabla P - \rho \nabla \Phi+ \frac{1}{4\pi} {\bf (\nabla \times B) \times B},
\end{equation}

\begin{equation}
   \rho \frac{D}{Dt}\left(\frac{e}{ \rho}\right) = -P \nabla \cdot {\bf v},
\end{equation}

\begin{equation}
{\partial{\bf B}\over\partial t} = {\bf\nabla\times}({\bf v\times B}),
\end{equation}
where $\rho$ is the mass density, $P$ is the total gas pressure plus radiation
pressure, 
${\bf v}$ is the fluid velocity, $e$ is the internal energy density,
$\Phi$ is the gravitational potential, and 
$\bf B$ is the magnetic field vector. We adopt an adiabatic
equation of state $P=(\gamma-1) e$ and consider models with
$\gamma=5/3$.

We perform simulations using the
pseudo-Newtonian potential of the central mass $\Phi_{PW} =
GM/(r-R_S)$, where $R_S = 2GM/c^2$ is the Schwarzschild radius,
introduced by Paczy\'{n}ski \& Wiita (1980).  
This potential approximates general relativistic effects
in the inner regions, for a nonrotating black hole.
In particular, the Paczy\'{n}ski--Wiita  potential
reproduces the last stable circular orbit at $r=3 R_S$
as well as  the marginally bound orbit at $r=2 R_S$.

Our calculations are performed in spherical polar coordinates
$(r,\theta,\phi)$. We assume axial symmetry about the rotational axis
of the accretion flow ($\theta=0^\circ$ and $180^\circ$).  The
computational domain occupies the radial range
$r_i~=~1.5~R_S \leq r \leq \ r_o~=~ 1200~R_S$, and the angular range
$0^\circ \leq \theta \leq 180^\circ$. The $r-\theta$ domain is
discretized on a non-uniform grid as in PB03b.

Our calculations use the ZEUS-2D code described by Stone \& Norman
(1992a,b).  We adopt PB03b's boundary and initial conditions but
we make two modifications to the initial conditions (see below).
An important element of the initial conditions is that the 
rotating gas has constant specific angular momentum, $l$, 
and is  confined in a wedge near the equator between 
$\theta=90^\circ -\theta_0$ and $\theta=90^\circ +\theta_0$.
The wedge of the rotating gas is truncated at the radius, $r(v_r=c_\infty)$
defined as the radius where
the initial radial velocity (as predicted by modified
Bondi velocity law, see Proga \& Begelman 2003a, hereafter PB03a) equals the sound speed at infinity, $c_\infty$.

The two modifications of the PB03b initial conditions are:
(i) we set the initial conditions exactly
as in PB03b and then reduce the density and internal energy
by a factor of 100
in the part of the computational domain where $v_\phi=0$ (outside
the rotating wedge) and (ii)
we consider two field geometries: 
a purely radial magnetic field as in PB03b and
a vertical magnetic field defined by the potential 
${\bf A} = (A_r=0, A_\theta=0, A_\phi= A r \sin\theta)$.
For $r \sin{\theta} > r(v_r=c_\infty)$, we 
scale the magnitude of the magnetic field
using a parameter, $\beta_o \equiv 8 \pi P_B(r_o)/B^2$
defined as the plasma parameter $\beta\equiv 8\pi P/ B^2$ at 
$r_o$, so that
$A =(2 \pi P_B(r_o)/\beta_o)^{1/2}$
(where $P_B$ is the gas pressure associated with the Bondi solution
at $r_o$).
For $r \sin{\theta} < r(v_r=c_\infty)$, we set the constant A to
a very small value. 

We introduce these two modifications to 
reduce the flow complexity. In particular,
we want the rotating flow to be dominant
and we want to follow the evolution of the rotating flow with 
as  little interference
as possible from other flow components.

We choose the following units: 
$r_0=R_S$,
$t_0=4\pi(R_S^3/2GM)^{1/2}$, 
$v_0=c$, 
$B_0=(4\pi\rho_\infty c^2)^{1/2}$, 
$\rho_0=\rho_\infty$
(the density at infinity for a classic  Bondi flow). The force
is in units of $f_0=c^4/4GM$ and the specific angular momentum
is in units of $l_0= 2 R_S c$.

\section{Results}

To simulate a simple inflow/outflow solution we must appropriately 
set the model parameters.
Our initial conditions help to promote 
the dominance the rotating flow and reduce effects
of the non-rotating flow. 
The importance of the non-rotating flow depends not 
only on the initial conditions
but also on the minimum value of the density allowed in 
the simulations, the so-called density floor, $\rho_f$.
We set $\rho_f=(1/r)^{1/2}$. 
For comparison, 
the maximum density at small radii achieved
during the simulation is about $10^4$.
To aid the rotating flow in reaching an organized solution
we set $l$  smaller
than the critical angular momentum, $2 R_S c$. [Otherwise the rotating gas will
form a rotationally supported torus which becomes turbulent
and generates a magnetized complex corona and outflow (e.g., PB03b).]
Here we describe results from the simulations with $l=0.8$.
We set $\theta_0=56^\circ$ (as in a fiducial model in PB03b).

For our choice of $l$, the flow near the equator is sub-Keplerian at all radii
and will likely stay  sub-Keplerian during the evolution with appreciable 
radial velocity (i.e., the equatorial flow may continue to accrete 
supersonically). This feature distinguishes our simulations from 
many previous ones where the focus was on the outflows from rotationally 
supported disks or torii. Our main focus is on formation of an outflow
from nearly radially falling gas away from the equator.

We have performed numerous simulations varying the magnitude and geometry
of the magnetic field, the specific angular momentum, the numerical resolution
and the density floor. 
Here we present results from two models: model A without a magnetic field
and model B for which $\beta_o=10^3$ and the initial magnetic field 
is vertical.
Model A is a reference model which illustrates the flow pattern when
accretion proceeds directly whereas
model B illustrates the dynamics and properties of
simplest accretion flow which generates outflow
without formation of rotationally supported disk and without development
of magnetorotational instability.

In the early phase of the evolution, when the flow
relaxes from the initial conditions, both models show the same behaviour. 
First, the gas near the equator falls in nearly radially onto  a BH. However,
the rotating gas closer to the poles diverges away from the equator
because of a lack of pressure equilibrium in the $\theta$ direction
due to the density and internal energy difference between
the equatorial wedge and the polar region. 
The flow at small radii becomes gradually radial,
with the density decreasing  between the
equator and the poles. After this early phase, the evolution of the 
two models proceeds differently. In particular, the flow in model A
settles down into a steady state of direct accretion whereas
in model B an outflow as well as a direct accretion flow form.

Figure~1 presents the flow pattern for model A at small radii 
at $t =3550$.
The figure shows the density map overplotted with
the direction of the poloidal velocity.
The flow has two components: 
(i) an equatorial inflow and 
(ii) a polar funnel inflow. The equatorial inflow has non-zero $l$
but its cicularization radius is inside the last stable orbit
therefore direct accretion occurs. The polar
funnel inflow has zero $l$. The density contrast between the two components
is due to the fact that at the outer boundary and for the initial conditions
the density of the non rotating gas is smaller than the density of the rotating
gas by a factor of 100. 
At $t=500$, the mass accretion rate, $\MDOT_a=0.08$
in units of the Bondi rate ($\MDOT_B$) and then gradually increases.
At the end of the simulation, $t=3550$, $\MDOT_a=0.123$ and still continues
to grow but very slowly. We estimate that $\MDOT_a$ will saturate
at the level of $\approx 0.125$. We stress that there is no indication of
an outflow in model A. This contrasts with the MHD counterpart
of this model.

In model B, the establishment of the equatorial inflow
is soon (at $t>715$) followed by a development of a bipolar outflow from  
the 'shoulders' of the inflowing gas. At first, the outflow is
confined to a very narrow range of $\theta$ but with time
this range increases. In particular, at t=5700,
the outflow is at $26^\circ \lesssim \theta \lesssim 48^\circ$ and
$132^\circ \lesssim \theta \lesssim 154^\circ$ at $r=20$. 
In other words, we observe that with time the boundary
between the equatorial inflow and the  bipolar outflow moves toward 
the equator. The mass loss rate associated with the outflow, 
$\MDOT_W$ increases with time. For example,
at $t=1500$, $\MDOT_W=0.0008$  while at $t=5700$, $\MDOT_W=0.008$
($\MDOT_w$ was measured at $r=175$). The late time evolution of $\MDOT_a$
for model B differs from that for model A. Namely,
for model B, $\MDOT_a= 0.08$ at $t=500$ 
and then gradually increases to 0.11 at $t=1500$. The mass accretion
rate stays at this level until $t=5000$ and then decreases
to 0.102 at $t=5700$.

The magnetic field evolves from vertical to nearly radial. 
This change in the field geometry is a natural consequence of
radially falling gas because the field lines are dragged in with
the gas. However, even at the end of the simulation,
the field did not evolve into a split monopole 
configuration due to the initial nonradial evolution of the flow.

Of course, the evolution of the field configuration
is accompanied by the growth of the field strength.
In particular, shear generates the toroidal field in the rotating flow.
The toroidal field is fastest growing component of the field in the
rotating gas and eventually dominates 
the poloidal field, $B_p=(B_r+B_\theta)^{1/2}$.
Additionally, 
the continuously growing field becomes dynamically important
(we find that an outflow forms in the regions where 
$\beta_\phi \equiv 8 \pi P/B_\phi^2<1$). 
In the polar region, the poloidal field is dominant,
$\beta_p \equiv 8 \pi P/B_p^2<1$) whereas
$\beta_\phi >1$).

Figure~2 shows the flow pattern at small radii at $t =5700$ for model B.
The left and right panels show density and $|B_\phi|$ maps, respectively.
The figure shows also the direction of the poloidal velocity and
poloidal magnetic field as well as an example of a streamline.
The flow has  four components: 
(i) a radial equatorial inflow,
(ii) a bipolar outflow
(the streamline shown in Fig. 2 is typical for this flow component), 
(iii) a polar funnel outflow, and
(iv) a polar funnel inflow.
The third and fourth components (we call collectively 
the funnel flow) are separated from the inflow/outflow solution
by the centrifugal barrier. We note that the funnel outflow
can be hardly distinguished from the bipolar outflow based
on the direction of the poloidal
velocity. Comparison of the panels of Fig. 2 
reveals  that across the boundary between the bipolar outflow
and the funnel flow, three significant changes  occur: 
(i) the density dramatically decreases; (ii)
$B_p$ changes suddenly direction, and (iii) $B_\phi$ increases.
We also note that the polar funnel inflow is confined to
a very narrow region along the rotational axis and its
properties/dynamics are difficult to capture by our numerical
approach because of the imposed density floor
and gas heating by artificial viscosity. These effects can
occasionally lead to production of a thermal outflow in the funnel 
(e.g., see the lower half of the left panel for $r\lesssim - 11$ 
near the rotational axis).
In the remaining part of the paper, we focus on 
the second component of the MHD flow -- the inflow/outflow solution.

The relative simplicity and slow time evolution of the inflow/
outflow solution helps us to  identify the forces responsible
for production of the outflow from the infalling gas.
For example, we have analyzed and compared all the terms in the equation of
motion along various streamlines.
The important force components in the radial direction are: 
gravity, $f_g$, gas pressure gradient, $f_p$, centrifugal force, $f_c$
and gradient of the  toroidal  field pressure, $f_m$ (the other
components of the Lorentz force are negligible).
In the latitudinal direction  $f_p$, $f_c$ and $f_m$ 
determine the total force, $f_t$.

Figure~3 presents various fluid and magnetic field
properties and the forces acting on the flow as a function
of the pathlength (the pathlength is measured from
$r=14$ and $\theta=53^\circ$)
along the typical streamline for the inflow/outflow solution.
To generate this plot, we assume that the solution
is steady. This assumption is justified
for our purposes here because the time changes in the flow
occur on the scales significantly longer than the time needed for the 
flow element to follow the segment of the streamline we analyze.

Fig.~3a shows that the motion in the radial direction is 
determined in the following way:
(1) the infall slows down and eventually stops
due to the centrifugal force. Up to the stagnation point, 
the fluid specific angular momentum stays nearly constant
(dash-dotted line in Fig. 3f).
The centrifugal force can balance gravity, despite
sub-Keplerian rotation on the equator, because $f_c$
increases from the equator toward the poles  
[$f_c \propto 1/(r^3 \sin^2{\theta})$ for $l=const$].
Near the stagnation point, two significant changes occur in the force
balance: (i) $f_m$ becomes significant and is directed
outward and (ii) $f_c$ increases compared to gravity because
the magnetic torque increases somewhat specific angular momentum of the fluid. 
The combined centrifugal and $B_\phi$ pressure 
forces overcome the combined gravity and gas pressure 
(the latter is directed inwards)
and accelerate the flow outward. Although 
$f_c > f_m$ in the outflow, $f_c$ alone
is not strong enough to accelerate the flow.
Therefore, the flow is not 'flung out' by the centrifugal
force as in Blandford \& Payne's (1982) magnetocentrifugal wind
but is  much more gradually pushed by the pressure of the toroidal field
(see Spruit 1996 and also below for a discussion of heavy loaded MHD winds).

Fig. 3b shows that as the infalling gas approaches 
the stagnation point, $f_p$ pushes the gas away from the equator against 
the centrifugal force. Near the stagnation point, two significant changes 
occur in the force balance in the $\theta$ direction: (i)  $f_m$ chances sign 
and (ii) $f_p$ increases so that it
alone can 'collimate' the gas during the early phase of the outflow
(this is done against not only $f_c$ and also against $f_m$).

Farther downstream from the stagnation point,  $f_m$ again changes sign
and becomes stronger than $f_p$. 
Thus, $f_m$ appears to play an important role in the outflow collimation 
at large radii whereas $f_p$ pushes the outflow
away from the equator during the initial acceleration. 
However, when the $\theta$ component of $f_m$ changes sign for 
the second time, $B_r$ changes sign too
(the dashed vertical lines in all panels of Fig. 3 mark the location
where $B_r$ changes sign downstream from the stagnation point). 
In ideal MHD, the magnetic field should be dragged in with the flow
and no changes in the field orientation should occur
within an organized outflow due to field freezing. However, 
in our solution of MHD equations using the grid based code,
the change of $B_r$ sign is due to
annihilation (reconnection) of oppositely
oriented fluxes (i.e., the polar funnel flux vs. the bipolar
outflow flux) on the grid-spacing scale.
We note this numerical limitation
and focus on the properties of the bipolar outflow 
upstream from the point where the polar funnel flux
annihilates the bipolar outflow flux.

The bipolar outflow is  unbound and it is heavy loaded.
To quantify the latter, we 
compute the so-called mass loading parameter
$\mu\equiv (v_p v_\phi/v^2_{Ap})^{1/2}$  
(where $v_{Ap}\equiv (B_p^2/4\pi\rho)^{1/2}$ 
is the poloidal Alfven spead) introduced by
Spruit (1996, see also eq. 2 in Anderson et al. 2004; note a difference
between the definiation of $\mu$ in the two references). 
At the stagnation point, $\mu=30$.
Fig.~3d shows that 
the toroidal field is higher than the poloidal field 
(compare solid, dotted, and dashed lines)
while Fig.~3f shows that
the outflow does not corotate with its base and the outflow
does not gain much of specific angular momemtum
(see  dashed and dash-dotted lines in Fig.~3f for
the angular velocity, $\Omega$ and specific angular momentum, respectively).
All these flow characteristics 
are consistent with the high value of $\mu$ at the stagnation point
(e.g., Spruit 1996).

Fig. 3e shows that the flow is super Alfvenic (dotted line).  
This figure also shows that
the ratio between the Poynting flux and kinetic energy flux, $F_P/F_K$
increases downstream from the stagnation point.
The energy flux ratio becomes of order of unity 
near the location of the bipolar outflow 'merging' with the funnel flow 
(i.e., the location mark by the vertical dashed line). 

\section{Summary and Concluding Remarks}

We present results from asymmetric HD and MHD simulations of 
a rotating gas collapsing onto a stationary BH. We consider
extreme cases where the gas rotation is so small that the circularization
radius is inside the last stable orbit.  Therefore one expects direct accretion
of all gas without formation of a rotationally supported torus
no mention of formation of an outflow.
After an initial transient, the HD flow, as expected, 
settles into a configuration where direct accretion
occurs and there is not outflow. This contrasts with the MHD flow,
which even for a very weak initial magnetic field, 
settles into a configuration with four
components: (i) a radial equatorial inflow,
(ii) a bipolar outflow, (iii) polar funnel outflow, and
(iv) polar funnel inflow.
Our focus here is on the second, unexpected component of the MHD flow.

The bipolar outflow is driven by 
the magnetic and centrifugal forces. Following the streamlines
of the bipolar outflow, we identify
four stages of the flow motion:
(i) a gravitational collapse, (ii) a gradual slow down and eventual
termination of the collapse by the centrifugal force,
(iii) launching of the outflow by gas pressure which
redirects the flow away from the equator and by the centrifugal force
which redirects the flow away from a BH in the radial direction 
(the latter force is enhanced owning to
a magnetic torque spinning up the gas), and (iv)
an acceleration by the centrifugal force and gradient
of the toroidal field pressure. 
The outflow is highly magnetized ($\beta< 1$) and 
the Poynting flux carries a significant fraction of the total
energy. 

Despite performing the simulation for thousands of orbits at the inner radius
the flow did not reach a steady state.
One indication of the time evolution is  that
an increasingly larger fraction of the equatorial inflow 
turns into the bipolar outflow. Additionally,  the
inflow very close to the equator can start turning back, 
the equatorial symmetry breaks, and
the flow near the equator starts to circulate. 
However, not all flow properties evolve with time.
For example, the angular velocity profile along the equator
is time steady and closely follows the  $1/r$ scaling
as expected for the constant $l$ flow.
Generally,
the flow evolves over a very long time scale 
into a state where the bipolar outflow 'sandwiches'
a smooth radial inflow or a complex flow near the equator 
(our test runs with $l=0.5$ show that
the equatorial flow can remain a smooth inflow for as long
as the simulations). We have observed a similar bipolar outflow in
almost all of our simulations
performed recently or performed in PB03b and Proga et al. (2003).
Our inflow/outflow solution resembles 
inflow/outflow self-similar solutions studied in the context of 
the core collapse leading to star formation 
(e.g., see the circulation region in Fig. 1 of Lery el. al 2002; 
see also Henriksen, R.N., \& Valls-Gabaud 1995,
and reference therein). We also note that our solution could be qualitatively
similar to inflow-outflow circulation found in the simulations
of the magnetized cloud contracting under self-gravity (Tomisaka 1998)
and to the funnel-wall jet found 
in fully relativistic numerical simulations of accretion disks in 
the Kerr metric (De Villiers t al. 2004). 

We finish with two remarks: (i) The inflow/outflow
solution is relevant to several astrophysical problems
where some of the magnetized collapsing gas may have very low
angular momentum (e.g., star forming regions, the Galactic center
and other supermassive BH environments where the BH is surrounded by
the diffused gas and mass losing stars, some GBRs and supernovae
where it is believed that the explosion is related to a collapse
of a massive rotating stellar envelope). (ii) Although
our physics, initial and boundary conditions are 
simple compared to those  in real systems,
our inflow/outflow solution is 
simple and probable robust, and will help to pinpoint the key elements
of real jets/outflows from low $l$ accretion flows.

ACKNOWLEDGMENTS: We thank Philp Armitage, Mitch Begelman, Roger Blandford, 
Scott Kenyon, Tom Gardiner, Bohdan Paczy\'{n}ski, 
Jim Stone, and Dimitri Uzdensky
for useful discussions. We also thank an anonymous referee for useful
comments that helped us clarify our presentation.
We acknowledge support from NASA under ATP grant NNG05GB68G
and support provided by NASA through grant  HST-AR-10305.05-A
from the Space Telescope Science Institute, which is operated
by the Association of Universities for Research in Astronomy, Inc.,
under NASA contract NAS5-26555.

\clearpage

\begin{figure}
\begin{picture}(180,200)
\put(380,-10){\includegraphics{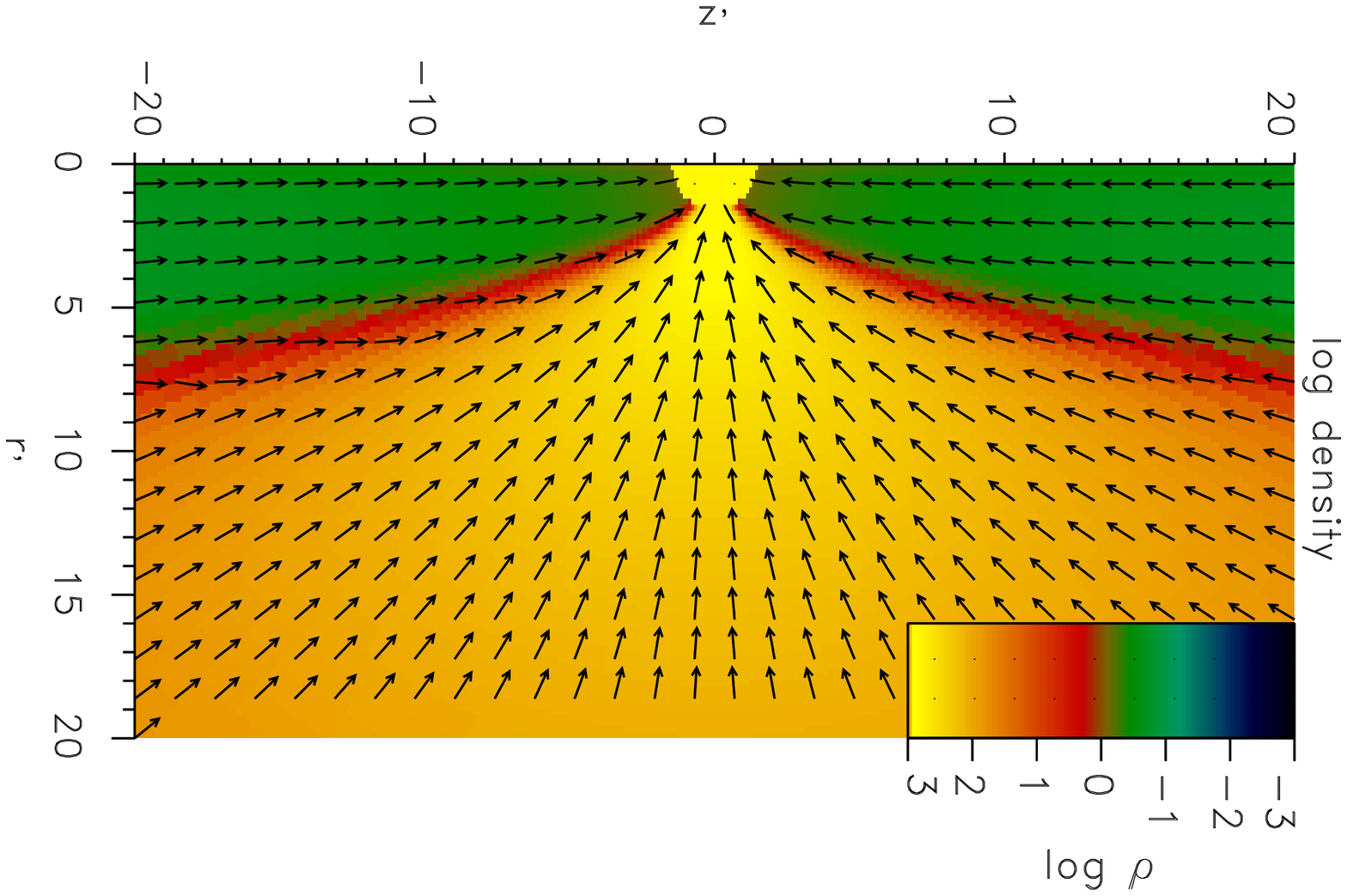}}
\end{picture}
\caption{
A map of logarithmic density overplotted with
the direction of the poloidal velocity for model A 
(a zero magnetic field case).}
\end{figure}

\clearpage

\begin{figure}
\begin{picture}(180,200)
\put(380,-10){\includegraphics{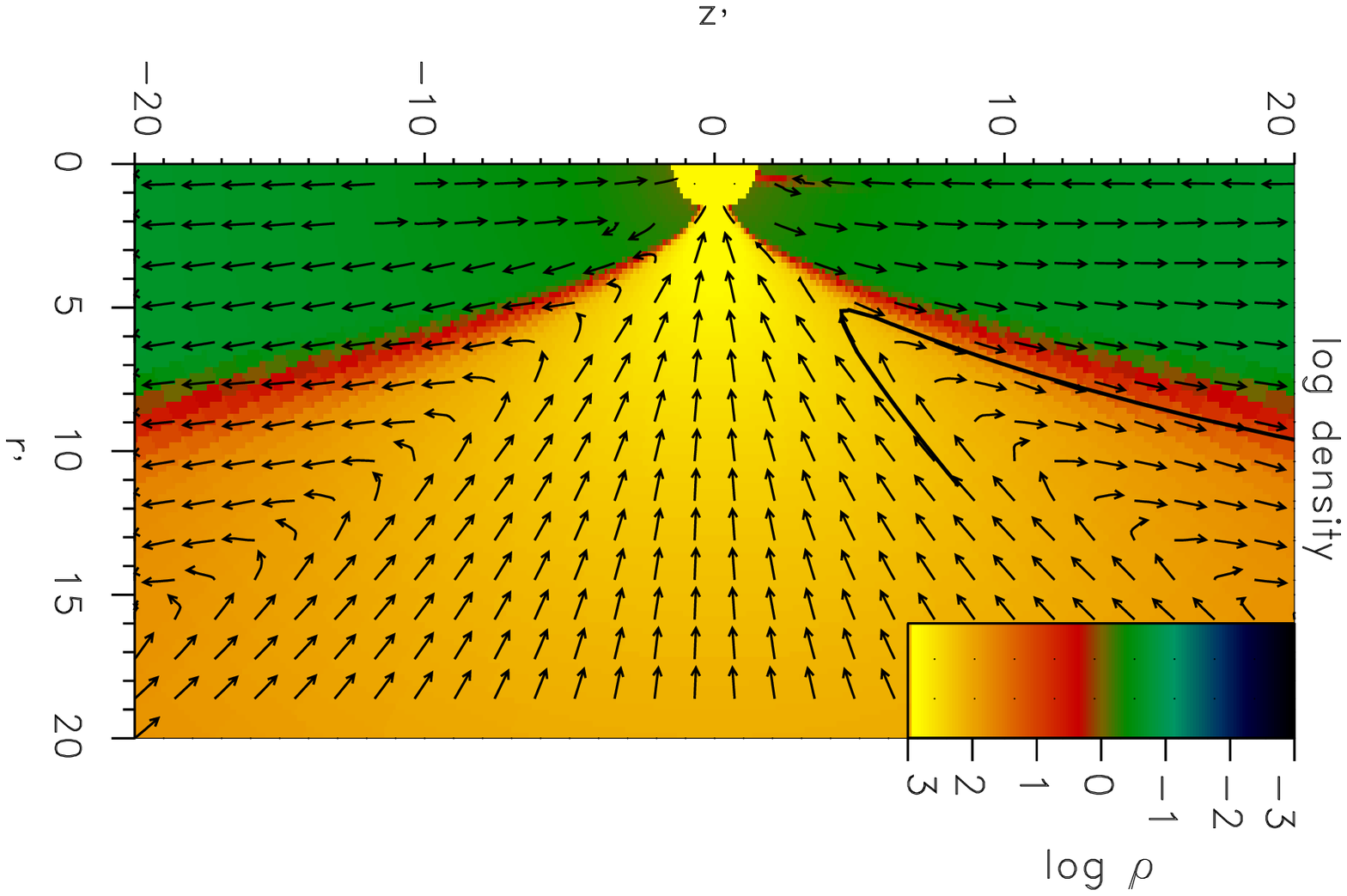}}
\put(630,-10){\includegraphics{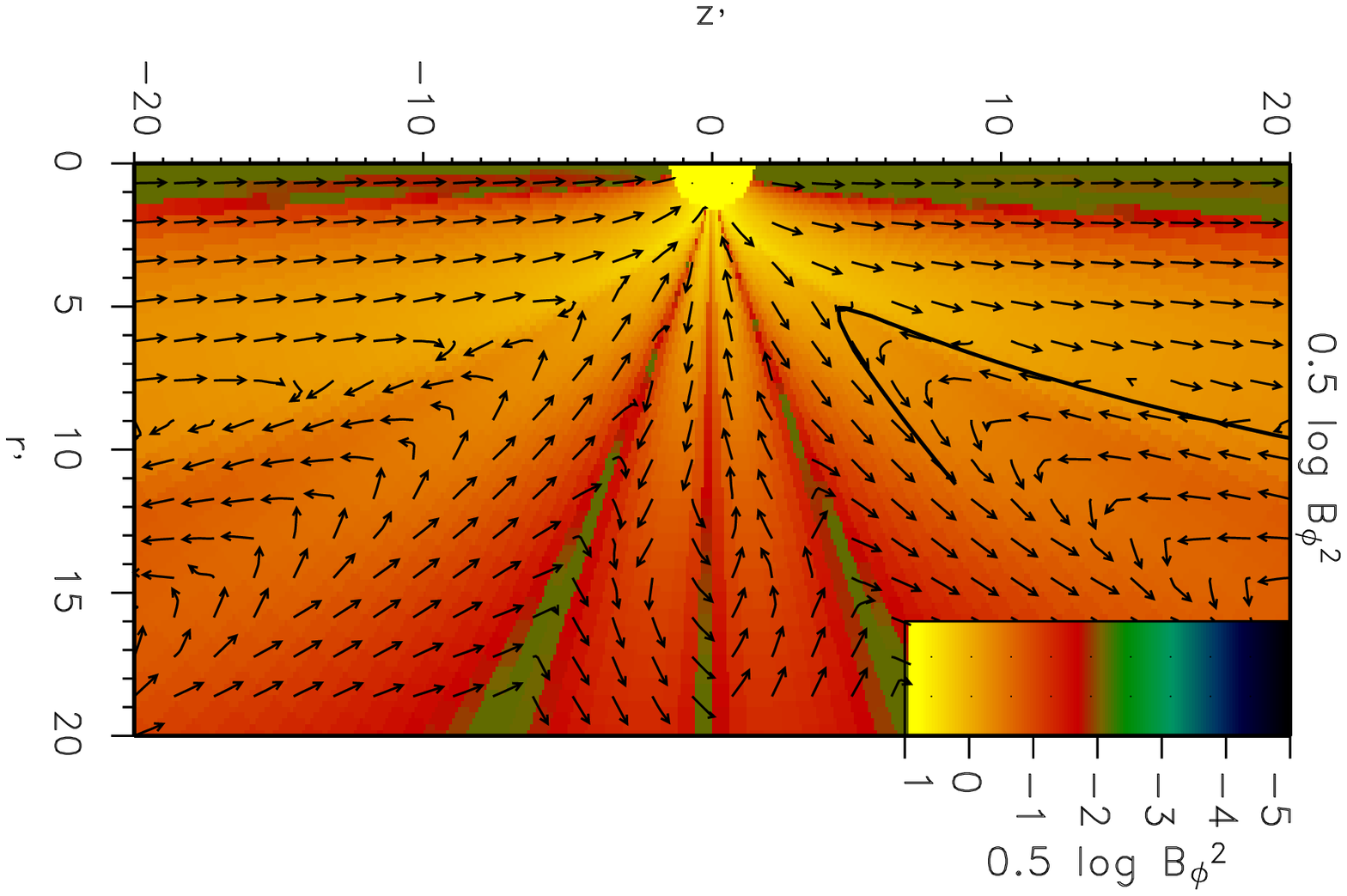}}
\end{picture}
\caption{
Maps of logarithmic density (left panel) and toroidal magnetic field 
(right panel) overplotted with an example of a streamline 
corresponding to an inflow/outflow for model~B 
(a non-zero magnetic field case). 
The maps are also overplotted with
the direction of the poloidal velocity and
the direction of the poloidal field 
(the left  and right panels, respectively).
}
\end{figure}

\begin{figure}
\begin{picture}(180,480)
\put(100,460){\includegraphics{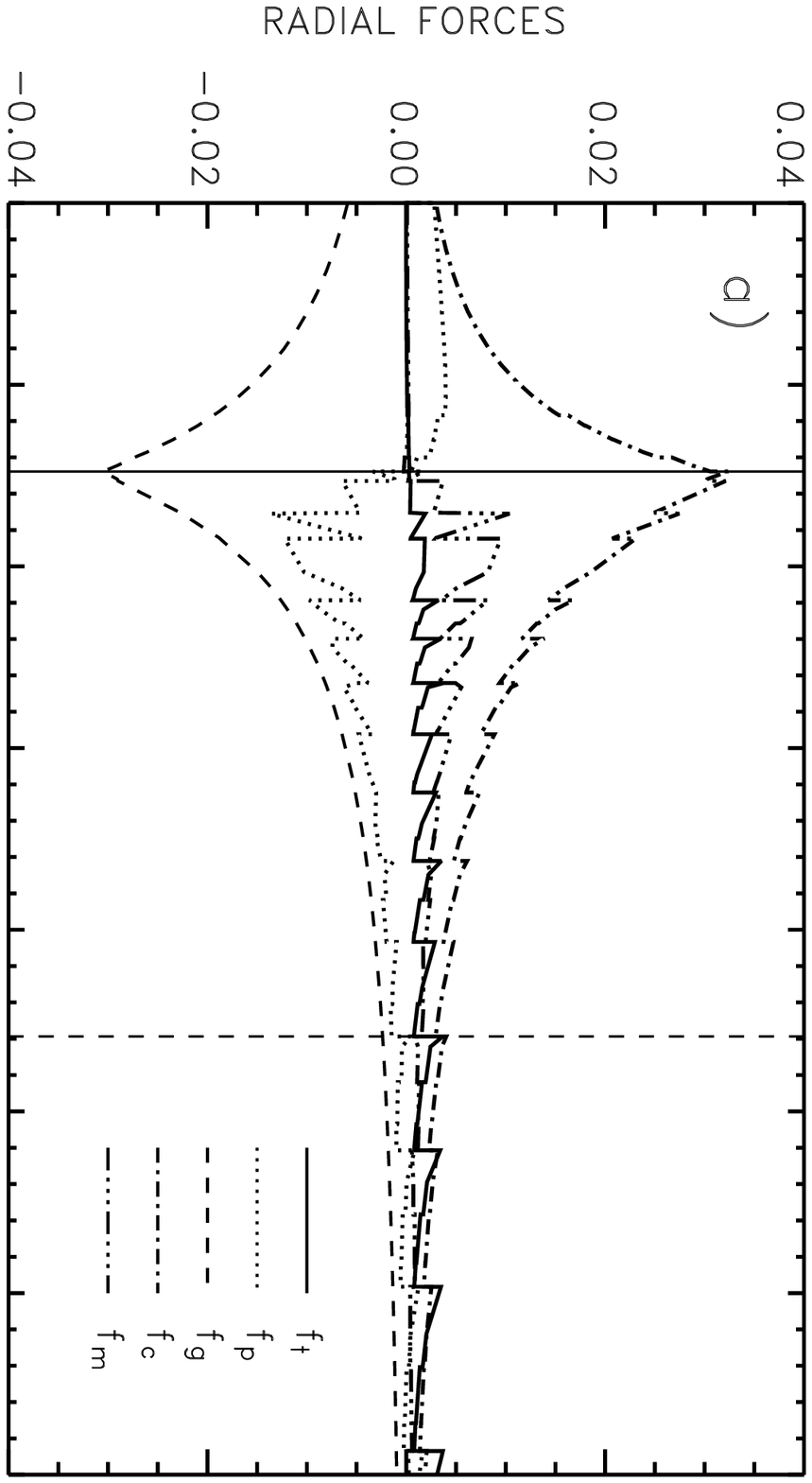}}
\put(100,340){\includegraphics{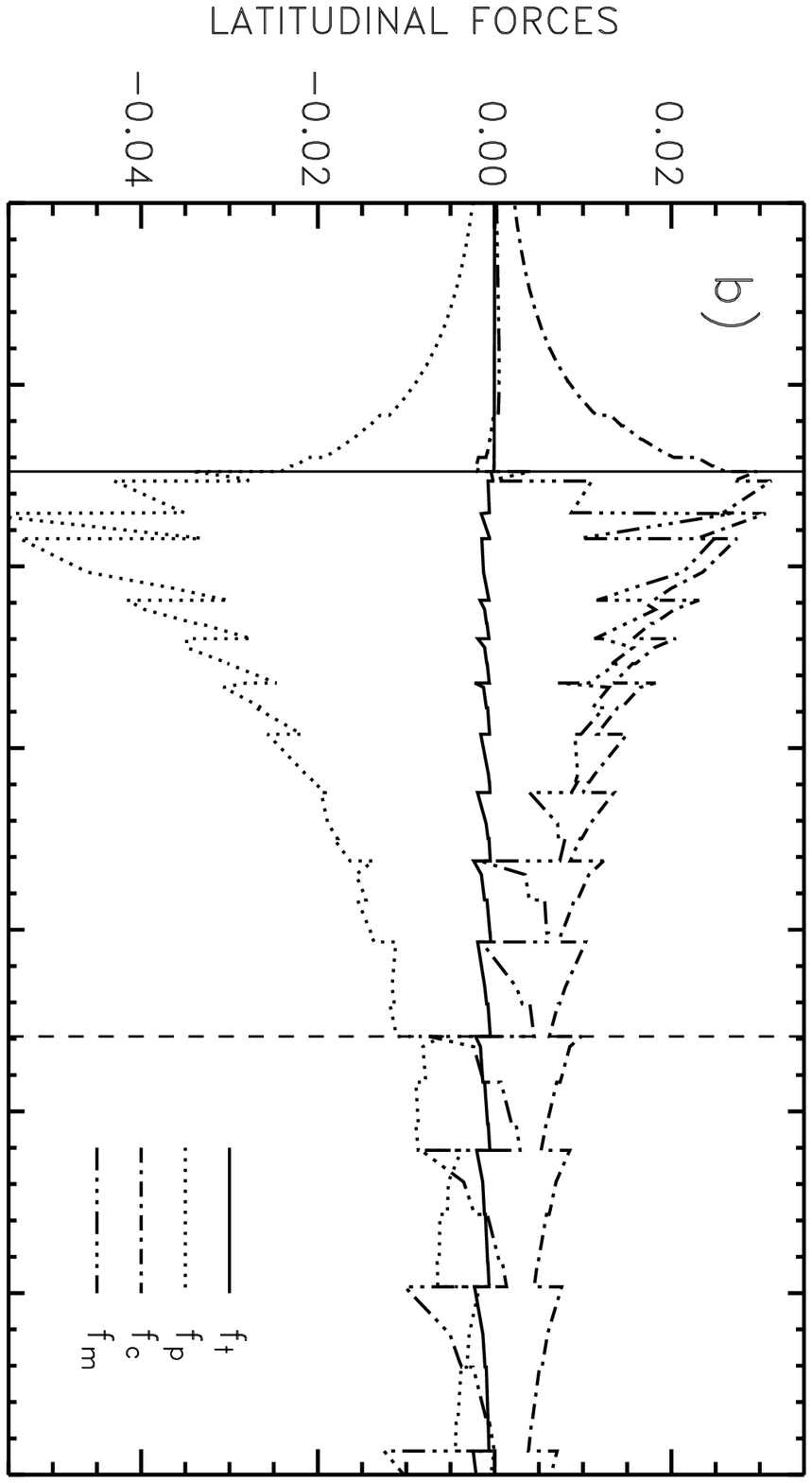}}
\put(100,220){\includegraphics{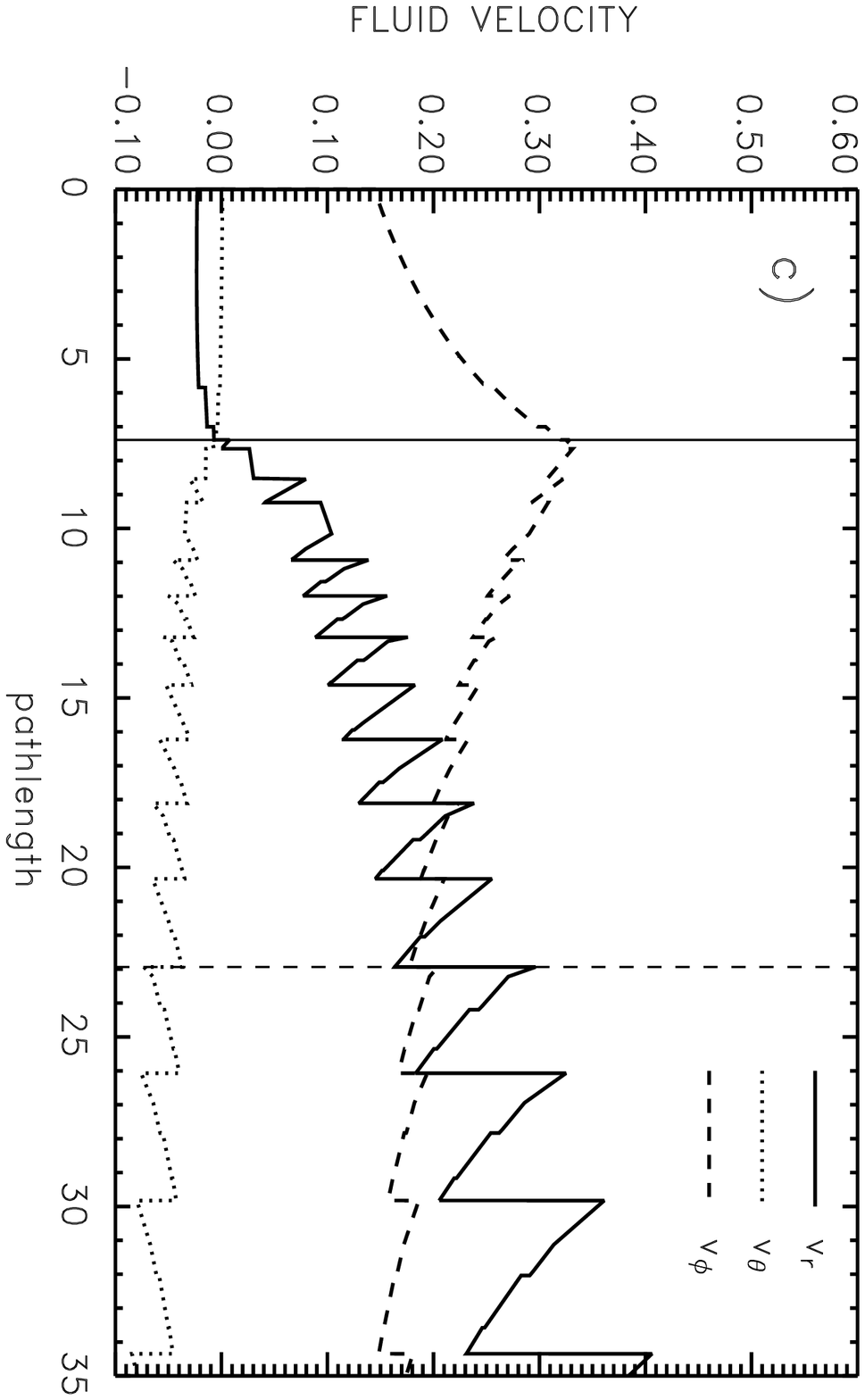}}

\put(340,460){\includegraphics{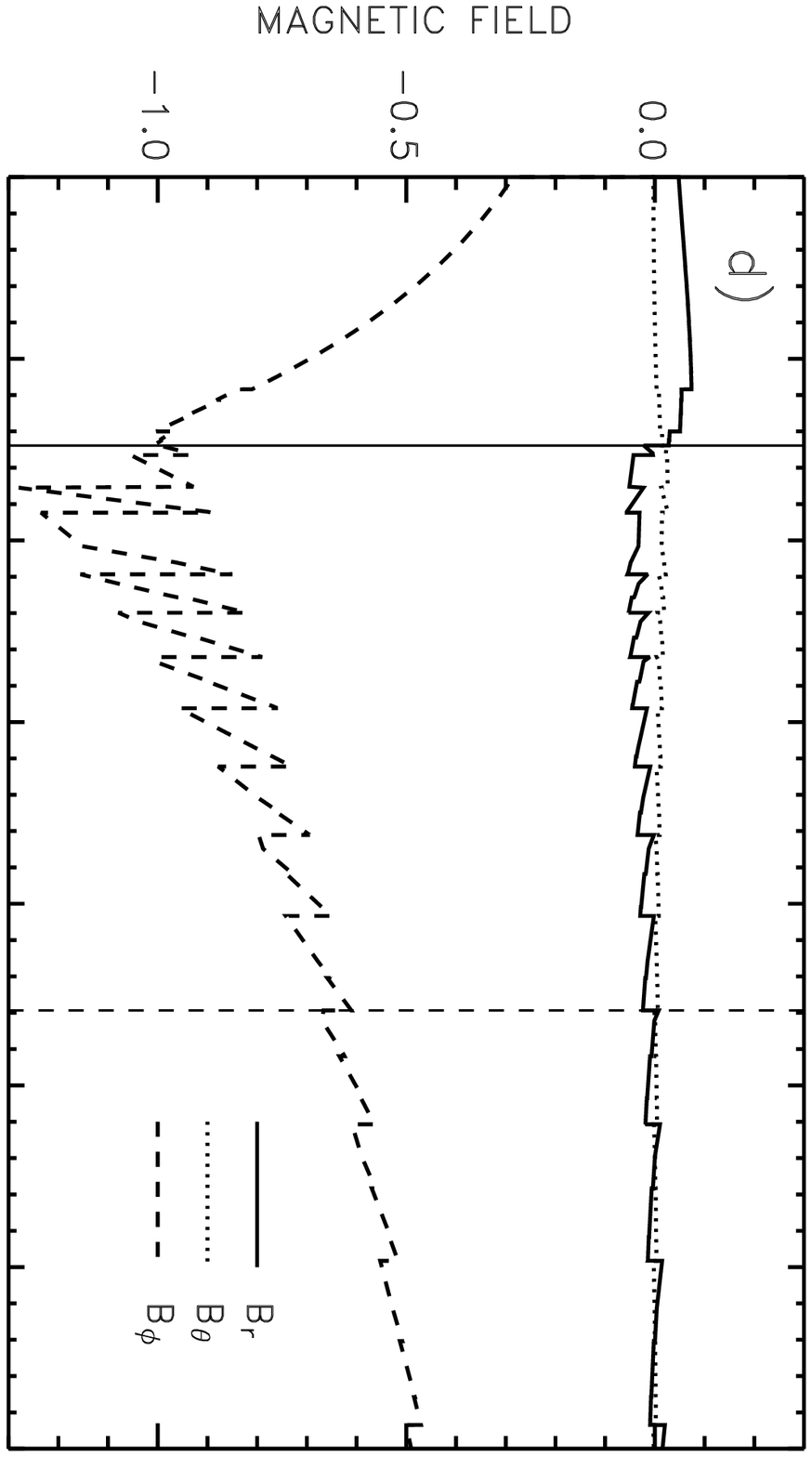}}
\put(340,340){\includegraphics{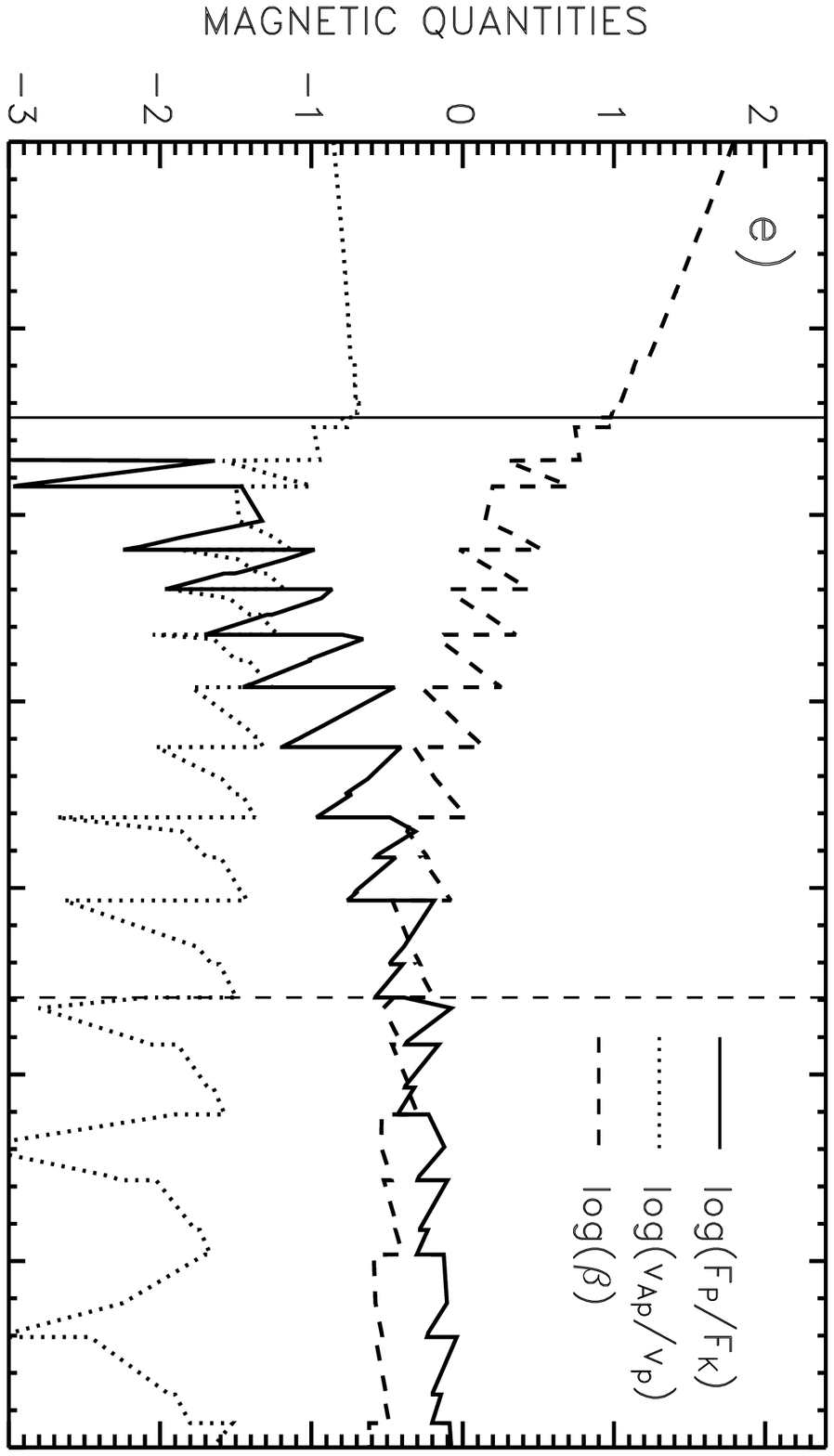}}
\put(340,220){\includegraphics{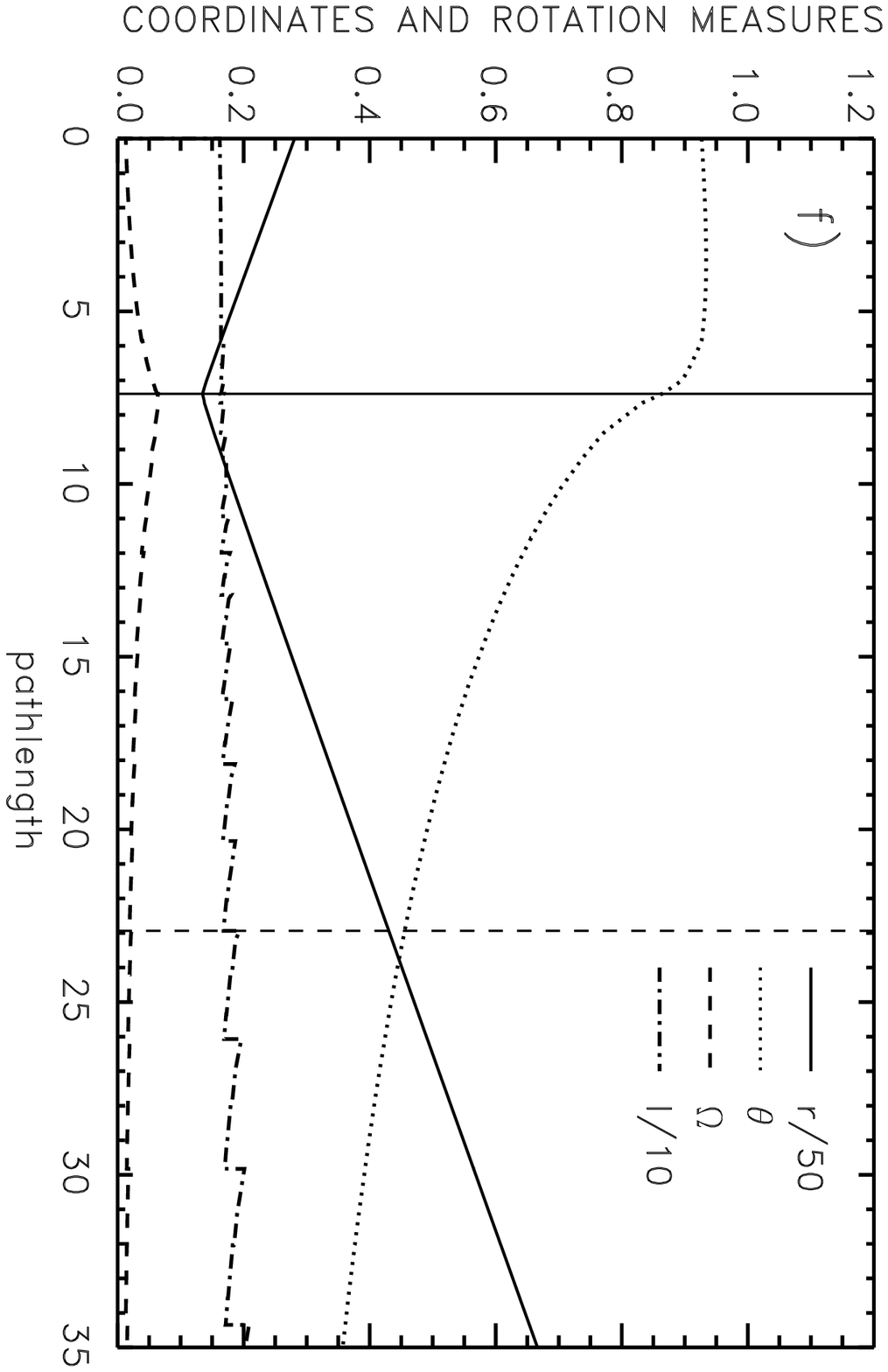}}

\end{picture}
\caption{\small{
{\it From top to bottom:}  The dominant forces
in the radial and latitudinal
directions (a and b panels, respectively)
along the streamline presented in Fig. 2.
The fluid and magnetic properties as a function of
the pathlength along the streamline (c-f panels).
For details see the labels and the main text.
Note that the specific angular momentum, $l$ is reduced
by a factor of 10 while the radius along the stream
line is reduced by a factor of 50
(dot-dashed line and solid line
in  panel f, respectively).
The solid  vertical lines in all panels correspond to the location where
the flow reaches the stagnation point 
(the minimum radius along the streamline).
The dashed  vertical lines in all panels correspond to the location where
the radial component of the magnetic field changes sign
and indicates the location where the magnetic field associated
with the bipolar outflow is advected into the grid cell
with a much stronger, oppositely  oriented magnetic field associated
with the polar funnel.
}}
\end{figure}

\end{document}